\documentclass[11pt]{article}
\usepackage[T1]{fontenc}
\usepackage{amsmath,amssymb,amsthm}
\usepackage[margin=1in]{geometry}
\usepackage{booktabs}
\usepackage{enumitem}
\usepackage[hidelinks]{hyperref}
\usepackage{xurl}
\newtheorem{theorem}{Theorem}
\newtheorem{proposition}[theorem]{Proposition}
\newtheorem{lemma}[theorem]{Lemma}
\theoremstyle{remark}
\newtheorem{remark}[theorem]{Remark}
\newcommand{\C}{\mathbb{C}}
\newcommand{\E}{\mathbb{E}}
\newcommand{\M}{\mathcal{M}}

\newcommand{\oai}{\cite{OAIFourier}}

\title{No complete set of mutually unbiased bases in $\C^6$\\ without Fourier vanishing}
\author{Jinze Yang$^{1}$, Yile Liu$^{2}$, Yangyang Li$^{3,*}$ and Xiu-Hao Deng$^{4,5,\dagger}$\\[1.5ex]
\small\itshape $^{1}$School of Physics, Xidian University, Xi'an 710071, China\\
\small\itshape $^{2}$School of Mathematical Sciences, Peking University, Beijing 100871, China\\
\small\itshape $^{3}$Key Laboratory of Intelligent Perception and Image Understanding of Ministry of Education of China,\\
\small\itshape Collaborative Innovation Center of Quantum Information of Shaanxi Province,\\
\small\itshape Institute of Interdisciplinary Quantum Science and Technology (IIQST),\\
\small\itshape School of Artificial Intelligence, Xidian University, Xi'an 710071, China\\
\small\itshape $^{4}$Shenzhen International Quantum Academy, Shenzhen 518048, China\\
\small\itshape $^{5}$Shenzhen Branch, Hefei National Laboratory, Shenzhen 518048, China\\[1ex]
\small $^{*}$\href{mailto:yyli@xidian.edu.cn}{yyli@xidian.edu.cn}\qquad
$^{\dagger}$\href{mailto:dengxiuhao@iqasz.cn}{dengxiuhao@iqasz.cn}}
\date{\today}
\begin{document}
\maketitle

\begin{abstract}
A recent exact certificate shows that $\C^6$ admits no seven mutually unbiased bases, hence at most five
\cite{OAIFourier}. That argument has two parts. The first proves the Fourier-vanishing conjecture of Matolcsi, Ruzsa
and Weiner for every complex Hadamard matrix of order six other than Tao's cubic matrix, and excludes the cubic matrix
separately. The second feeds this vanishing into a positivity certificate on mixed moments of a hypothetical complete
family. We show that the first part is not needed. We keep the same six families of test functions and the same linear
identities, but we never set a moment to zero because of Fourier vanishing. New positive definite integer matrices
still give the contradiction. The only inputs are orthonormality, unbiasedness, the second-moment (2-design)
identity of a complete family, and the trivial bound $|\text{moment}|\le 1$. A standard-library program checks the
certificate in exact integer and rational arithmetic in under a minute and a half. We also record computational
observations that are not part of the proof: a systematic $S_6$-symmetric family of test functions appears to close
the argument with relations of total order at most~$7$ instead of~$11$, and at order~$9$ two orbit types of test
charges appear to suffice. In the prime-power dimensions $4,5,7,8$, where complete families exist, the same pipeline
finds no certificate, as it must.
\end{abstract}

\section{Introduction}

Two orthonormal bases of $\C^d$ are \emph{mutually unbiased} if every inner product between a vector of one and a
vector of the other has modulus $d^{-1/2}$. At most $d+1$ bases of $\C^d$ can be pairwise mutually unbiased, and
complete families of $d+1$ such bases exist whenever $d$ is a prime power
\cite{Ivanovic1981,WoottersFields1989}. Write $N(d)$ for the maximal size of a mutually unbiased family in $\C^d$.
Dimension six is the smallest case outside the prime-power constructions. Zauner conjectured that $N(6)=3$
\cite{Zauner2011}. Weiner proved that $d$ mutually unbiased bases in $\C^d$ always extend to a complete family
\cite{Weiner2013}. Hence $N(6)\le 5$ is equivalent to the non-existence of seven mutually unbiased bases in $\C^6$.

Several computer-assisted proofs addressing $N(6)$ appeared in the autumn of 2026. OpenAI's companion preprint
\cite{OAIMUB} proves $N(6)=3$ by a discretisation of the space of complex Hadamard matrices of order six. Its validity
is stated under hypotheses on binary64 floating-point arithmetic and the compiler that ran the computation. Jaswal
carried out an independent certified computation of that discretisation proof, again under a binary64-correctness
hypothesis \cite{Jaswal2026a}. Cárdenes Wuttig and Tindall classify all triples of mutually unbiased bases in $\C^6$ up
to equivalence and show that none extends to a fourth basis; their proof is certified with interval arithmetic.
They also give an analytic proof of the Fourier-vanishing conjecture discussed below \cite{CWT2026}. Jaswal gives a further computer-assisted classification of triples and recovers that there are no four
mutually unbiased bases \cite{Jaswal2026b}.

The preprint \oai\ proves the weaker bound $N(6)\le 5$ with a certificate that uses \emph{only} integer and rational
arithmetic and has no floating-point hypothesis at all. The bound $N(6)\le 5$ is therefore not the news here. What is
at stake is the structure of the exact certificate. That proof has two parts.
\begin{enumerate}[label=(\Alph*)]
\item Sections~3--5 of \oai\ prove the Fourier-vanishing conjecture of Matolcsi, Ruzsa and Weiner
  \cite{MRW2013}, which arose in the Fourier-analytic approach to mutually unbiased bases \cite{Matolcsi2012}: $g_H(\pi\alpha)=0$ for $\alpha=(1,1,1,-1,-1,-1)$ and every coordinate permutation
  $\pi$, for every complex Hadamard matrix $H$ of order six that is not equivalent to Tao's cubic matrix
  \cite{Tao2004}. The steps are an exact single-matrix identity \cite[(10)]{OAIFourier}, the dichotomy
  \cite[Corollary~3.1]{OAIFourier}, a classification of the cubic alternative \cite[Section~4]{OAIFourier}, and an
  exclusion of the cubic matrix from complete families \cite[Proposition~5.2]{OAIFourier}.
\item Section~6 of \oai\ is a positivity certificate on mixed moments of a hypothetical complete family. It uses the
  vanishing as an input \cite[(27)]{OAIFourier}.
\end{enumerate}

The point of this note is that part~(A) is not needed for the bound. Using the framework of
\cite[Sections~6--7]{OAIFourier} with the vanishing switched off, the obstruction is already visible from identities
that hold in every dimension for every complete family. These are orthogonality, unbiasedness and the second-moment
identity. The single-matrix theorem of \oai\ remains a result of independent interest. We only show that the
complete-family exclusion does not depend on it.

\section{Setting and the identities used}

We use the notation and the probability space of \cite[Section~6.1]{OAIFourier}. Bare equation numbers such as
(12) or (27) refer to the numbering of \oai; this note has no numbered equations of its own. Suppose that $B_1,\dots,B_7$ are
mutually unbiased orthonormal bases of $\C^6$. Choose a reference basis uniformly, a uniform permutation of its
vectors, and a uniform bijection from the labels $0,\dots,5$ to the remaining bases. Let $H_s$ be the transition
matrices and $g_s=g_{H_s}$ the normalised character sums \cite[(1)]{OAIFourier}. Let $\M(\mathbf s;\mathbf a)$ be the
mixed moments \cite[(28)]{OAIFourier} and $m(A)$ the ordered-pair injective moments \cite[(29)]{OAIFourier}. All of
these are real numbers of modulus at most one \cite[(30)]{OAIFourier}. For a uniformly sampled column $z$ of $H_0$, the
test functions are
\[
  F_0(a)=g_0(a)z^{-a},\qquad F_1(a)=\tfrac15\textstyle\sum_{s=1}^5 g_s(a)z^{-a}
  \qquad\text{\cite[(31)]{OAIFourier}.}
\]

\paragraph{Identities used.} We use exactly the following valid relations.
\begin{enumerate}[label=(R\arabic*)]
\item The contraction relations of \cite[Lemma~3.2]{OAIFourier}, eqs.~(5) and~(6). These come from orthogonality of a
  single transition matrix. We use the formal seven-row relations~(6) generated from exponent matrices with seven
  formal rows, two to four columns and total order $7$ to $11$, enumerated as in \cite[Section~7.2, (i)--(iii)]{OAIFourier}.
\item The vanishing of every moment that has a charge of order one. This is row orthogonality, the first half of
  \cite[(12)]{OAIFourier}, and the first half of \cite[Lemma~6.2(ii)]{OAIFourier}. When such a moment has a single
  label, its pair expansion is recorded as a homogeneous equation, as in \cite[Section~7.3(ii)]{OAIFourier}.
\item The singleton reduction \cite[Lemma~6.2(iii), (34)]{OAIFourier}. It follows from the second-moment identity
  $\sum_{s\ne r}g_s(a)=0$ for charges of order two \cite[(12)]{OAIFourier}. This identity is the projective 2-design
  property of a complete family \cite{KlappeneckerRoetteler2005}. It is the only identity used that requires the
  family to be complete.
\item The symmetries \cite[Lemma~6.2(i)]{OAIFourier} and the single-label expansion
  \cite[Lemma~6.2(iv), (35)]{OAIFourier}. The latter is the equality-pattern expansion of \cite[Lemma~3.3]{OAIFourier}.
\item The mixed identities \cite[(36), (37)]{OAIFourier}, obtained from column orthogonality and unbiasedness,
  for all charge triples of total order at most $11$, enumerated as in \cite[Section~7.3]{OAIFourier}.
\item The Gram formulas \cite[Lemma~6.1]{OAIFourier} and the bounds \cite[(30)]{OAIFourier}.
\end{enumerate}

\paragraph{Identities omitted.} In contrast with \cite[(27) and Lemma~6.2(ii)]{OAIFourier}, \emph{no moment is set to
zero because one of its charges is a permutation of $\alpha$}. Neither the identity \cite[(10)]{OAIFourier}, nor the
dichotomy \cite[Corollary~3.1]{OAIFourier}, nor the classification of the cubic case
\cite[Section~4]{OAIFourier}, nor the cubic-pair obstruction \cite[Proposition~5.2]{OAIFourier} is used.

\begin{lemma}\label{lem:valid}
For every set of seven mutually unbiased bases of $\C^6$, the relations (R1)--(R6) hold for the actual moment
vector, whether or not the transition matrices satisfy \cite[(27)]{OAIFourier}.
\end{lemma}

\begin{proof}
\cite[Lemma~6.2]{OAIFourier} is stated under (27). Its proof, however, invokes (27) only for the second half of
part~(ii). Parts (i), (iii), (iv) and (v) and the order-one half of (ii) are proved there from uniform sampling,
from (12), from the equality-pattern expansion \cite[Lemma~3.3]{OAIFourier}, and from (13) and (14).
\cite[Lemma~3.2]{OAIFourier} concerns one Hadamard matrix and uses only its orthogonality. \cite[Lemma~6.1]{OAIFourier}
uses only the sampling. The bounds (30) hold because each moment averages products of unimodular numbers or of
averages of them.
\end{proof}

\section{The certificate}

\begin{theorem}\label{thm:main}
There do not exist seven mutually unbiased bases in $\C^6$. Hence $N(6)\le 5$.
\end{theorem}

The theorem follows from Weiner's theorem \cite{Weiner2013} and the following finite statement.

\begin{proposition}\label{prop:cert}
Let $(\lambda,\mathrm{ws})$ run over the six lists of \cite[Section~6.3]{OAIFourier}, unchanged, and let
$Y_\lambda$ be the Young-symmetrised vectors of test functions \cite[(39)]{OAIFourier}. Let $Z_\lambda$ be the
symmetric integer matrices in the accompanying data file, given by their lower triangles as in
\cite[(38)]{OAIFourier}. Then:
\begin{enumerate}[label=(\alph*)]
\item each $Z_\lambda$ is positive definite;
\item using only (R1)--(R6), for every complete family,
\[
  5\sum_\lambda \E\bigl(Y_\lambda^* Z_\lambda Y_\lambda\bigr) \;=\; f+\sum_{\rho\in R}\beta_\rho X_\rho ,
\]
where each $X_\rho$ is one of the original moments $m(A)$ or $\M(\mathbf s;\mathbf a)$, and
\[
  f=-\frac{821605927536823}{38880}\approx -2.1132\cdot 10^{10},\qquad
  \sum_{\rho\in R}|\beta_\rho|=\frac{5531688869329}{12960}\approx 4.2683\cdot 10^{8}.
\]
\end{enumerate}
\end{proposition}

\begin{proof}[Proof (computer-assisted, exact)]
The computation follows \cite[Section~7]{OAIFourier} step by step, with the three changes listed below.
\begin{enumerate}
\item The pair relations (R1) are generated and reduced by the contraction~(5). Exact substitution rules are then
  certified by \cite[Lemma~7.1]{OAIFourier}: modular elimination proposes the rules, and an exact integer check
  $WP=0$ on the witness rows proves them.
\item The mixed relations (R5) are generated for every charge triple of total order at most~$11$, after the reductions
  (R2)--(R4), and certified in the same way.
\item Each $Y_\lambda$ is expanded by \cite[(39)]{OAIFourier}, and each Gram entry is replaced using
  \cite[Lemma~6.1]{OAIFourier}. The factor~$5$ is the one in \cite[(40)]{OAIFourier}. The reductions and the certified
  substitutions are then applied.
\end{enumerate}
The changes relative to \oai\ are these. The zero rule for permutations of $\alpha$ is switched off. The coordinates
are ordered by the default sort rather than by the priority lists of \cite[Section~7.1]{OAIFourier}; this choice
changes which coordinates are free, but adds no relation. The matrices $Z_\lambda$ are new. As noted in
\cite[Section~7.1]{OAIFourier}, the value of $f+\sum_\rho|\beta_\rho|$ depends on the coordinate order, and only its
sign for the order actually used matters. The matrices $Z_\lambda$ here were computed for the default order, which the
verifier fixes deterministically, so no priority list has to be supplied.

The pair table has $8871$ nonzero equations, $909$ coordinates and $897$ witness rows, as in \oai. The mixed table
has $40618$ equations, $3583$ coordinates and $2842$ witness rows; with vanishing, \oai\ has $46041$, $3429$ and $2767$.
After substitution, $481$ non-constant symbols remain: $10$ pair moments, $320$ two-label and $151$ three-label mixed
moments. All of them are irreducible original moments, and $21$ of the mixed ones contain a permuted~$\alpha$.
Under~(27) these $21$ would vanish; here they are bounded by~$1$ like all the others.

Positive definiteness is checked by exact rational Schur complements, as in \cite[Section~7.4]{OAIFourier}. All $98$
pivots are positive. The smallest pivot of each matrix lies between $0.97$ and $11.6$, for entries of size up
to about $10^9$. The check is exact, so this small margin is immaterial. Real Gram forms with a positive definite
real matrix are nonnegative on complex vectors. The bounds (30) then give
\[
  0\;\le\; 5\sum_\lambda \E\bigl(Y_\lambda^* Z_\lambda Y_\lambda\bigr)\;\le\; f+\sum_\rho|\beta_\rho|
  \;=\;-\frac{201252715232209}{9720}\;<\;0,
\]
a contradiction. Every arithmetic step that certifies an equality or an inequality is integer or rational.
\end{proof}

The relative margin is large: $\sum_\rho|\beta_\rho|/|f|\approx 0.020$, against $\approx 0.293$ for the certificate of
\oai, which uses vanishing. Both ratios are invariant under rescaling of the matrices $Z_\lambda$.

\section{Verification}\label{sec:verif}

The verifier \texttt{verify\_d6.py} and the data file \texttt{cert\_d6\_data.json} accompany this note. Their SHA-256
checksums are
\begin{center}\small
\texttt{8149c17e766767712fa6a1724a9d9f22154ce2f473df8b38a652aad19d86df19} \ (\texttt{verify\_d6.py})\\
\texttt{2f7b602b28799008f78a28744ae6c0e0ee02633038b0f21bf8707a4c7080f3e7} \ (\texttt{cert\_d6\_data.json})
\end{center}
The verifier uses only the Python standard library (Python $\ge 3.8$) and regenerates every relation from scratch.
The command is \texttt{python3 -I verify\_d6.py cert\_d6\_data.json}. The internal checks are assertions, so the
program refuses to run when assertions are disabled (\texttt{-O} or \texttt{PYTHONOPTIMIZE}); the final sign test
$f+\sum_\rho|\beta_\rho|<0$ is an explicit check. On one core of a Linux workstation (Intel Core i7-13700KF) the run
takes $45$\,s with CPython~3.10.12 and uses about $0.38$\,GB of memory. It prints the rank tables and the values of
Proposition~\ref{prop:cert}. An earlier revision of the verifier, without these two safeguards,
printed the same output on a Linux cluster node with CPython~3.9.18 ($82$\,s) and CPython~3.12.4 ($69$\,s).

The program is an adaptation of the verifier printed in \cite[Appendix~A]{OAIFourier}. After normalising formatting,
$34$ of its functions are unchanged. These include the canonical forms, the contraction, the modular elimination and
rational reconstruction, the exact substitution check, the Young-symmetrised test functions and the Gram tagging. The
mathematical differences are exactly the following. The test for a permuted~$\alpha$ always returns false. The
single-matrix check of \cite[(10)]{OAIFourier} is absent. No priority lists are used. The pivot test is $p>0$, the
exact criterion for positive definiteness, instead of the margin $p>40000$ of \oai. A key that never occurs in a
relation is kept as a free symbol instead of raising an error. This is sound, because such a symbol is still an actual
moment bounded by~$1$, and with the present data it never happens.

As a cross-check, an independent C++ re-implementation of the relation generation and exact elimination
reproduces exactly the pair table, the mixed table of \oai\ with vanishing together with its published values of $f$
and $\sum|\beta|$, and the mixed table without vanishing used here. That program has not been run on the matrices
$Z_\lambda$ of this note. The proof rests on \texttt{verify\_d6.py} alone.

\section{Remarks}

\begin{remark}[Lower order; computational observation]\label{rem:lower}
The six lists of \oai\ need relations of total order~$11$. With the vanishing switched off, our pipeline finds no
certificate for them when all relations are cut at total order $9$ or $10$, nor when only the pair relations are cut at
order~$8$. The rationalised bounds are then positive: $+74.6$, $+77.5$ and $+3.66$.

A systematic family does better. Take Young-symmetrised test functions
$y_\mu F_t(v)$ for all partitions $\mu$ of $6$, with $y_\mu$ acting on all six coordinates, $t\in\{0,1\}$, and $v$
running over balanced charges of order at most~$3$. Optionally split each block into its real and imaginary parts
using $Y(-a)=\overline{Y(a)}$. With all relations of total order at most~$9$ (pair table $197/55/45$, mixed table
$5363/465/335$), our pipeline produced exact rational upper bounds of $-8.54$, and of $-9.57$ with the split. A variant
that keeps the relations as equality constraints with rational multipliers, instead of eliminating them, gave
$-9.97$. Test functions symmetrised only over the five non-reference coordinates gave $-13.24$.

In each case the floating-point dual of a semidefinite program was rationalised, and the resulting bound was then
evaluated exactly. These certificates have not yet been packaged as standalone data with a verifier, and they have not
been independently checked. We therefore record them as a computational observation, not as part of the proof. With
test charges of order at most~$2$ the corresponding semidefinite program was numerically feasible, so no certificate
of this type was found at that order.
\end{remark}

\begin{remark}[Prime-power dimensions]\label{rem:sanity}
The relation generator is written for a general dimension $D$. It uses $D+1$ formal rows in (R1), the factor
$1/(D+1-l)$ in (34), and the Gram weights $D-1$ and $D-2$. We ran the same pipeline with order-$3$ systematic test
functions, in several symmetrisation variants, and relations of total order at most $11$ for $D=4,5,7,8$. In these dimensions complete families exist, so
no certificate can exist. In every case the exact bound was positive, so the pipeline found no certificate, as it must.
The positive values ranged from $6\cdot 10^{-8}$ ($D=4$) to $1.5\cdot 10^{3}$ ($D=7$).

In several of these runs the semidefinite solver reported a slightly negative optimum, for example
$-1.2\cdot10^{-6}$ for $D=5$. Only the exact evaluation decides. This is a consistency check of the generator and of
the exact evaluation, not evidence about the strength of the method.
\end{remark}

\subsection{Further computational observations (\texorpdfstring{$d=6$}{d=6})}\label{sec:further}

The following refines Remark~\ref{rem:lower}. Its status is that of a computational observation, not of a theorem.
Each item was obtained with our independent C++ implementation of the relation generator (the program of the
cross-check in Section~\ref{sec:verif}). The floating-point dual of a semidefinite program was rationalised, and the
bound was then evaluated in exact rational arithmetic, in the variant of Remark~\ref{rem:lower} that keeps the
relations as equality constraints with rational multipliers. The rationalised Gram matrices are checked to be
positive semidefinite by exact elimination, and the bound $f+\sum_\rho|\beta_\rho|$ is computed exactly; a negative
value is a contradiction, provided that the generated relations are correct. The values depend on the normalisation of
the matrices, so only their signs matter. None of these certificates has been packaged as standalone data with a
verifier or checked independently, and none is used in the proof of Theorem~\ref{thm:main}.

The test functions are those of Remark~\ref{rem:lower}: $y_\mu F_t(v)$ for all partitions $\mu$ of $6$, $t\in\{0,1\}$,
and balanced charges $v$ of order at most~$3$, with each block split into real and imaginary parts. Up to permutation
and negation, these charges fall into $11$ orbit types. ``Relation order'' is the total order up to which
(R1)--(R5) are generated. The vanishing~(27) is never used.
\begin{enumerate}[label=(\alph*)]
\item \emph{Relation order $7$ suffices.} With all relations of total order at most~$7$ ($243$ relation rows), the
  exact bound is $-1.70<0$; at order~$8$ it is $-1.87$.
\item \emph{Two orbit types suffice at relation order $9$.} At relation order~$9$, if all test charges of
  type $(2,1,-1,-1,-1,0)$, or all of type $(2,1,-2,-1,0,0)$, are removed, the pipeline finds no certificate; if any
  one of the other nine types is removed, it still finds one. The charges of these two types alone, with
  $t\in\{0,1\}$ and all $\mu$, give the exact bounds $-0.227$ at relation order~$9$ and $-0.943$ at order~$11$. At
  relation order~$7$ the pipeline finds no certificate for these two types alone, whereas it does for the full family
  by~(a).
\item \emph{The single-matrix relations are not needed.} Omitting from (R1) all formal seven-row relations~(6) of
  \cite[Lemma~3.2]{OAIFourier} still gives certificates, with exact bounds $-9.88$ at relation order~$9$ and
  $-33.5$ at order~$11$.
\item \emph{The orbit of $\alpha$ is not needed.} Removing all test charges that are permutations of
  $\alpha=(1,1,1,-1,-1,-1)$ still gives certificates, with exact bounds $-9.98$ at relation order~$9$ and $-28.98$ at
  order~$11$. Conversely, for test charges of order at most~$2$ together with the orbit of $\alpha$, the pipeline
  finds no certificate at relation order $9$ or $11$.
\end{enumerate}

\section*{Data availability}
The verifier \texttt{verify\_d6.py}, the certificate data \texttt{cert\_d6\_data.json}, the checksums and the logs
of the reference runs are included as ancillary files of the arXiv version of this note. The verifier is an
adaptation of the verifier of \cite{OAIFourier}, which is distributed under the Apache License~2.0; it is released
under the Apache License~2.0, with the modifications listed in Section~\ref{sec:verif}. The certificate data are
released under the Creative Commons Attribution~4.0 International licence (CC~BY~4.0).

\section*{Use of AI}
Claude Opus~5.5 (Anthropic) was used to write and run the code (the adaptation of the verifier, the relation
generator and the semidefinite-programming pipeline). The authors take full responsibility for the content.

\end{document}